\documentclass[
]{ceurart}

\usepackage{listings}
\begin{document}

\copyrightyear{2024}
\copyrightclause{Copyright for this paper by its authors.
  Use permitted under Creative Commons License Attribution 4.0
  International (CC BY 4.0).}

\conference{CHIRP 2024: Transforming HCI Research in the Philippines Workshop, May 09, 2024, Binan, Laguna}

\title{How Fitts' Fits in 3D: A Tangible Twist on Spatial Tasks}


\author[]{Faith {Griffin}}[%
email=faith_griffin@dlsu.edu.ph,%
]
\author[]{Kevin {Abelgas}}[%
email=kevin_abelgas@dlsu.edu.ph,%
]
\author[]{Kriz Royce {Tahimic}}[%
email=kriz_tahimic@dlsu.edu.ph,%
]
\author[]{Andrei Kevin {Chua}}[%
email=andrei_chua@dlsu.edu.ph,%
]
\author[]{Jordan Aiko {Deja}}[%
email=jordan.deja@dlsu.edu.ph,%
]
\author[]{Tyrone Justin {Sta. Maria}}[%
email=tyrone_stamaria@dlsu.edu.ph,%
]

\address[1]{De La Salle University, Manila, Philippines}

\begin{abstract}
Expanding Fitts' Law into a 3D context, we analyze PointARs, a mixed reality system that teaches pointer skills through an object manipulation task. Nine distinct configurations, varying in object sizes and distances, were explored to evaluate task complexity using metrics such as completion time, error rate, and throughput. Our results support Fitts' Law, showing that increased distances generally increase task difficulty. However, contrary to its predictions, larger objects also led to higher complexity, possibly due to the system's limitations in tracking them. Based on these findings, we suggest using tangible cubes between 1.5" and 2" in size and limiting the distance between objects to 2" for optimal interaction in the system's 3D space. Future research should explore additional configurations and shapes to further validate Fitts' Law in the context of 3D object manipulation in systems like PointARs. This could help refine guidelines for designing mixed reality interfaces.
\end{abstract}


\maketitle

\section{Introduction and Background}

\par
\begin{figure}
    \centering
    \includegraphics[width=1\linewidth]{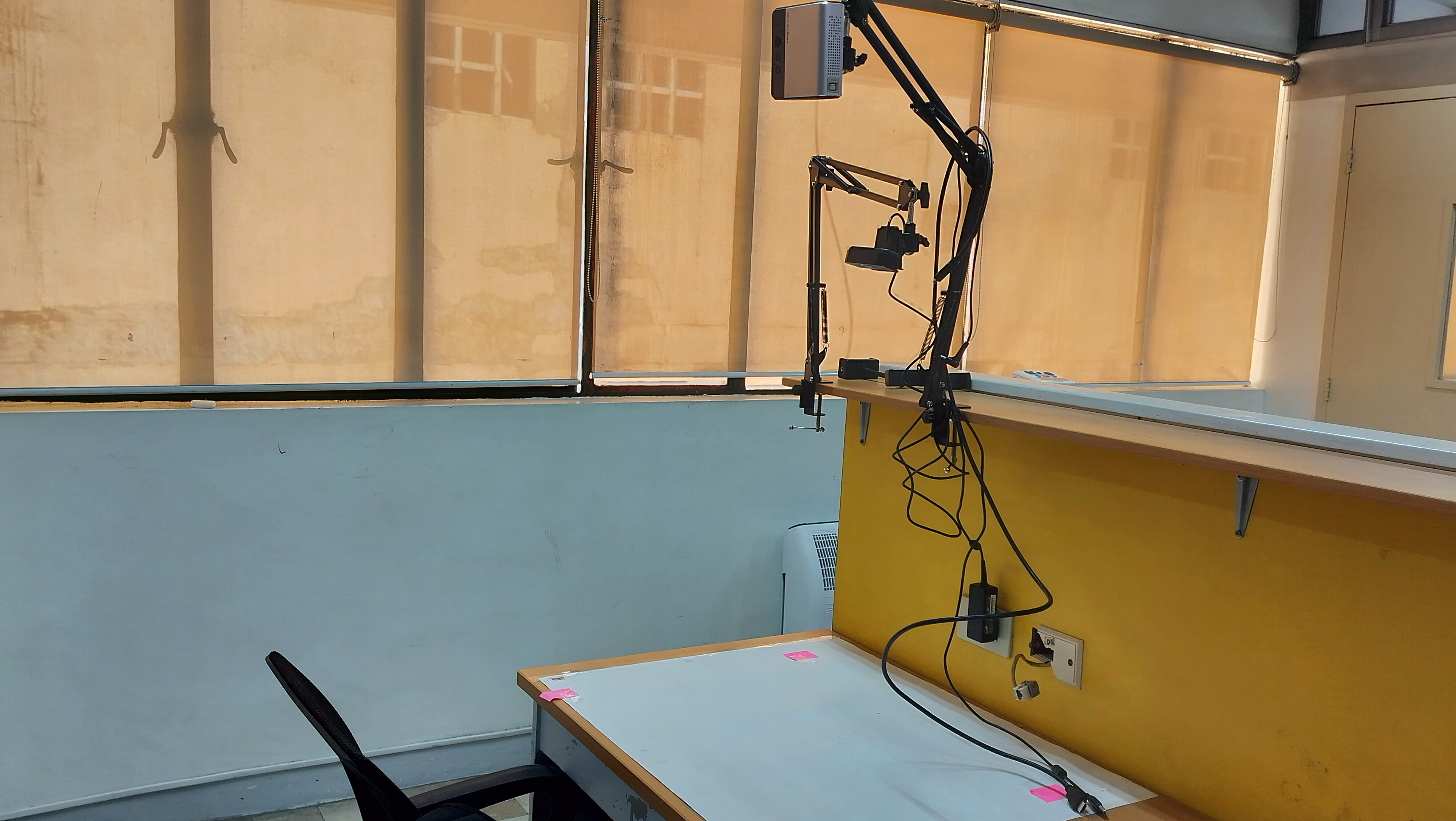}
    \caption{Current actual space setup of the PointARs prototype.}
    \label{fig:pointars-prototype}
\end{figure}

\begin{figure}
    \centering
    \includegraphics[width=0.5\linewidth]{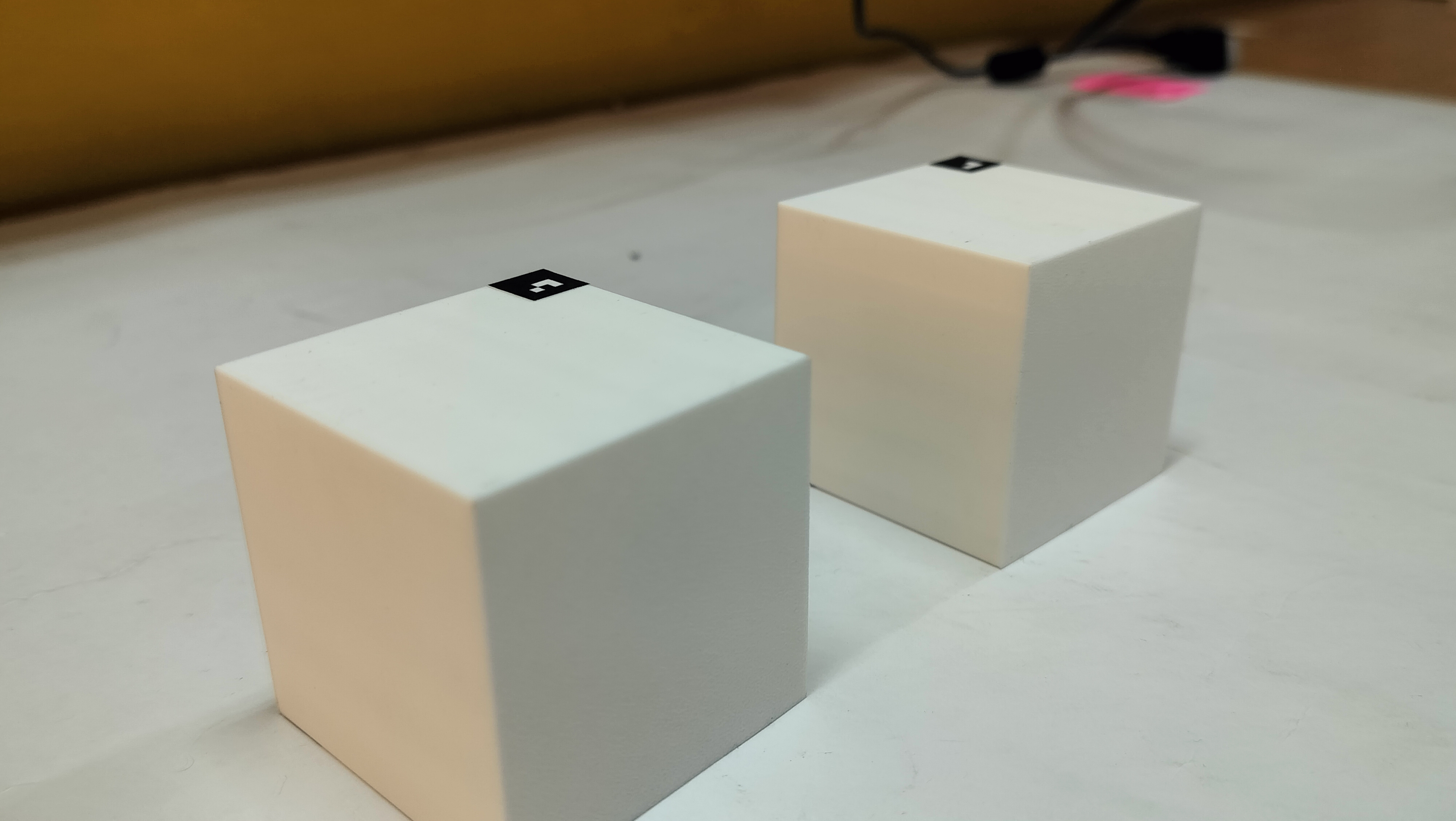}
    \caption{A photo of some of the tangibles in the current PointARs system.}
    \label{fig:tangible-variable}
\end{figure}

 Learning pointers can be challenging, especially for novice programmers, yet mastering them is essential for beginners in computer science. We present PointARs, a Mixed Reality Training System (MRTS) that combines Mixed Reality (MR) and tangible user interfaces (TUI) to help novices understand pointers (see Figure \ref{fig:pointars-prototype}). In this study, we build upon the existing design of PointARs, using it as a foundation for iterative improvements and empirical evaluations. Specifically, we focus on redesigning the system's tangible cubes, particularly their current 2x2x2 inch dimensions (shown in Figure \ref{fig:tangible-variable}), by conducting a series of target acquisition tests. Our aim is to inform future comparative studies assessing the effectiveness of PointARs as a complement to traditional learning methods, such as coding demonstrations and video presentations.

\par Tangible interactions have been explored in educational settings, offering several advantages. These interactions can enhance engagement and foster reflection by combining physical actions with digital augmentations \cite{marshall2007tangible}. Additionally, they support embodied cognition and help develop 3D mental visualizations \cite{yannier2016bridging}. Tangible systems are considered naturally intuitive, lowering barriers to participation, especially for novices \cite{marshall2007tangible}. The challenges novice programmers face in understanding pointers can be addressed using MR TUIs, which enhance learning by providing an interactive environment that allows learners to physically manipulate objects, making the experience more tangible and intuitive \cite{resnyansky2018potential}. MR TUIs can also simplify the concept of pointers, making them easier to understand \cite{de2016teaching}.

\par
\begin{figure}
    \centering
    \includegraphics[width=1\linewidth]{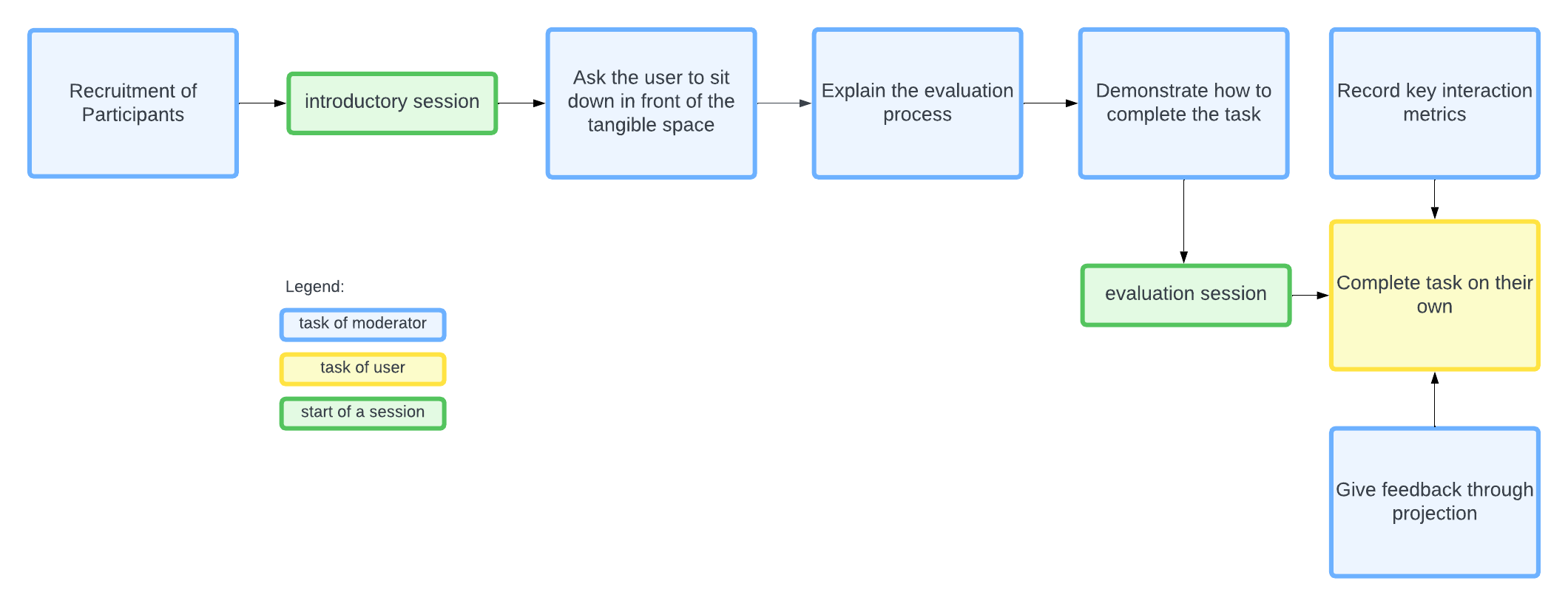}
    \caption{Overview of the steps in the evaluation protocol.}
    \label{fig:eval-protocol}
\end{figure}

 This study seeks to enhance 3D object manipulation within PointARs by applying Fitts' Law to examine how the size and arrangement of tangible cubes affect interaction metrics like completion time, error rate, and throughput (TP). Figure \ref{fig:eval-protocol} provides an overview of our evaluation pipeline, which will be detailed in the following sections, followed by a discussion of our findings and conclusions.

\section{Proposed Improvements and Rationale}
\par This study aims to expand the application of Fitts' Law to 3D object manipulation, inspired by the methodologies of Ha and Woo (\citeyear{ha2010empirical}), who examined various virtual hand techniques in a Tangible Augmented Reality (TAR) environment. We focus on how the size and spatial arrangement of tangible cubes affect key interaction metrics such as completion time, error rate, and throughput within PointARs’ tangible space.

\par The primary task for participants involves bumping a tangible cube to a stationary target cube located at varying distances from the center of the projection space. This mirrors the 3D object manipulation task in Ha and Woo’s study \cite{ha2010empirical}, adapted to the context of PointARs’ tangible space. By varying the size of the cubes and the distance between them, we aim to analyze the effect of these factors on the identified interaction metrics. We measured completion time in milliseconds (ms), from the moment the user grabs a cube until the task is completed, and error rate in millimeters (mm), using the Euclidean distance from the central position of the stationary target cube to the movable cube.

\par Our objectives are to (a) assess how the size of the cubes affects the interaction metrics, (b) evaluate the impact of varying distances between the cubes, (c) analyze the effect of different pairings of cubes with varying sizes, and (d) refine the cube dimensions based on these findings to enhance user interaction. Based on these objectives, we hypothesize that: (a) As the size of the cubes increases, completion time and error rate will increase, and throughput will decrease; (b) Greater distances between cubes will increase completion time and error rate while decreasing throughput; (c) Pairs of cubes with varying sizes will increase completion time and error rate and decrease throughput; and (d) Optimizing cube size and spacing will result in reduced reaction times, indicating more efficient interaction. Potential confounding variables include user familiarity with handling cube-shaped objects and hand size.

\section{Methodology}
\subsection{Description of the Tangible Fitt’s Law Test Program}
\par To measure completion time, we integrated an object-grabbing detection application into PointARs' existing Unity-based system. This application detects when a participant grabs a tangible cube, allowing us to measure the completion time. It uses MediaPipe Hands for hand detection and tracking, Ultralytics' YOLO (specifically the yolov5s model) for real-time object detection, and OpenCV for video processing. The system detects interactions by checking if a participant’s fingertips are near a tangible cube's center. Once a grabbing action is detected, the application displays a "Grab Detected!" message on the frame and sends a "grabbed" signal to a predefined server address and port via UDP sockets.

\par We also implemented a DataReceiver script within PointARs’ system, working alongside the grabbing detection program. This script uses UDP networking to receive messages from the grabbing detection system, enabling real-time adjustments to PointARs' tangible space based on user interactions. It manages a sequence of five trials per test, where participants adjust tangible cubes to match target positions. Progress is tracked through user input and the proximity between objects. Real-time feedback on task duration is displayed in the projection space, and task completion triggers a green-colored background. The script records interaction metrics such as completion time and error rate, storing the data in a CSV file for subsequent analysis.



\subsection{Data Collection}
\begin{figure}
    \centering
    \includegraphics[width=1\linewidth]{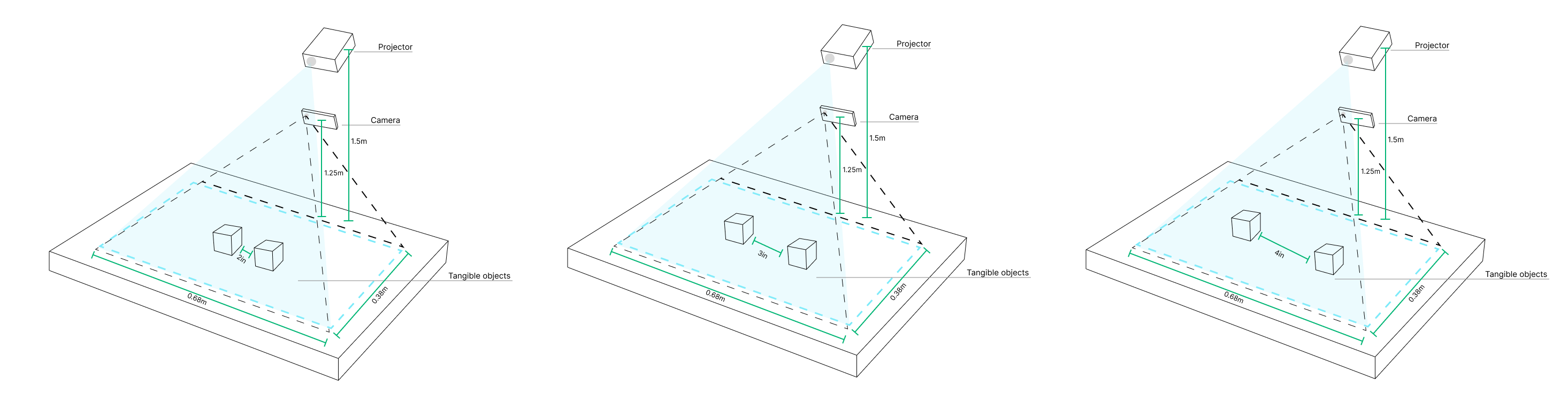}
    \caption{Tangible space schematics of the experiment setup during data collection}
    \label{fig:project-schematics}
\end{figure}

\par Using a purposive convenience sampling method, we involved three undergraduate computer science students from De La Salle University Manila as participants. They interacted with tangible cubes of different sizes (1.5x1.5x1.5, 2x2x2, 2.5x2.5x2.5 inches), placed at varying distances (2, 3, and 4 inches apart) from each other, as shown in Figure \ref{fig:project-schematics}. The cubes were paired based on size in the following combinations: 2” \& 2”, 1.5” \& 2”, 2” \& 1.5”, 2.5” \& 2”, 2” \& 2.5”, 1.5” \& 2.5”, 2.5” \& 1.5”, 2.5” \& 2.5”, and 1.5” \& 1.5”, where the first size represents the stationary cube and the second the movable cube.

\par Data collection was conducted using PointARs' actual setup, recording the completion time and error rate for each participant as they completed the designated tasks. A total of nine tests were performed per participant, consisting of five trials for each of the three tasks. Each test involved a different pairing of cubes based on their dimensions and varying distances between them. A task was considered complete once the movable cube made contact with the stationary cube and projected feedback was displayed. Before the evaluation, participants were given an explanation and demonstration of the task.



\section{Results and Analysis}
\begin{figure}
    \centering
    \includegraphics[width=1\linewidth]{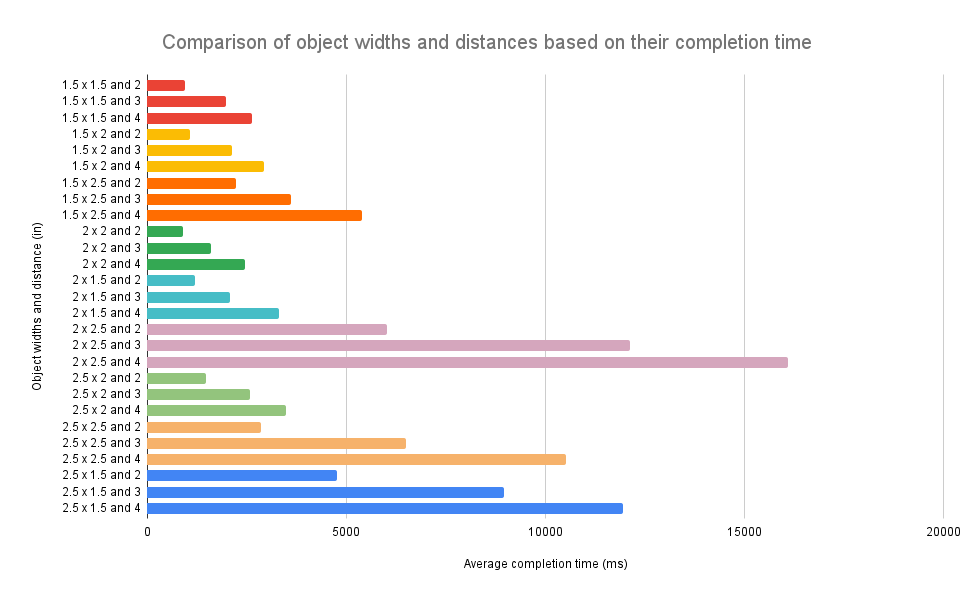}
    \caption{Average completion times per object size and distance.}
    \label{fig:completion-time}
\end{figure}

\begin{figure}
    \centering
    \includegraphics[width=1\linewidth]{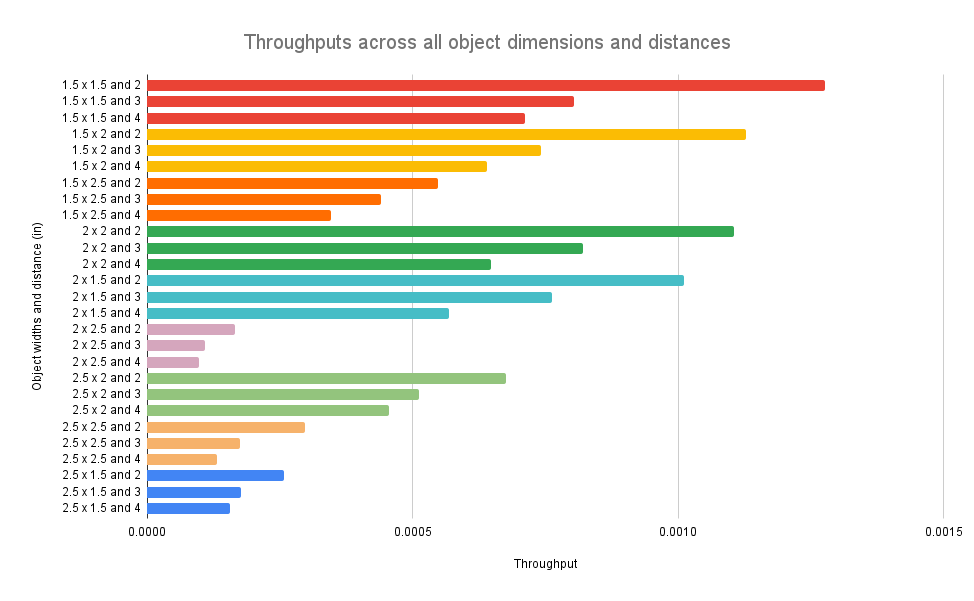}
    \caption{Throughputs per object size and distance.}
    \label{fig:tp}
\end{figure}

\begin{figure}
    \centering
    \includegraphics[width=1\linewidth]{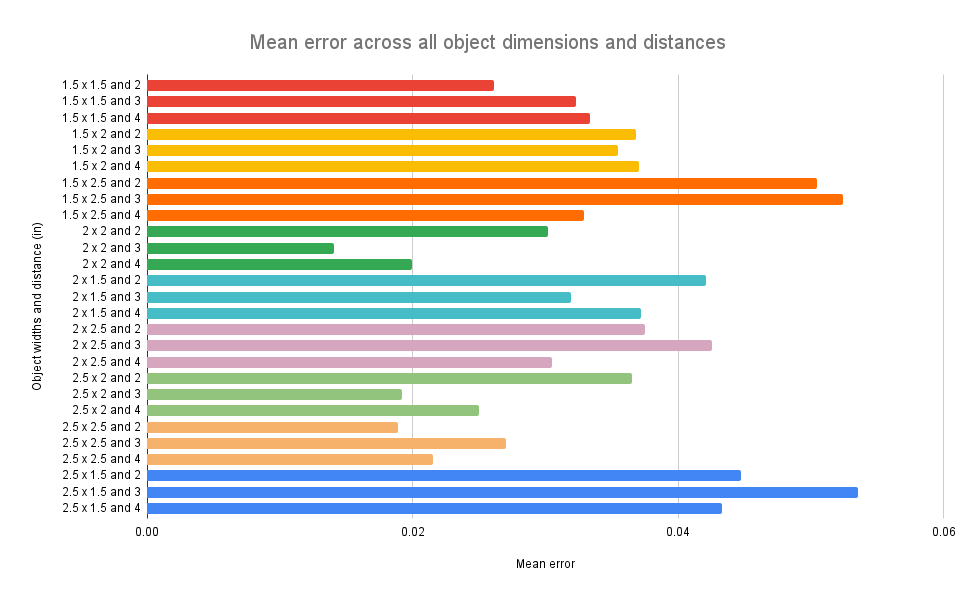}
    \caption{Error Rate per object size and distance.}
    \label{fig:error}
\end{figure}

\begin{figure}
    \centering
    \includegraphics[width=1\linewidth]{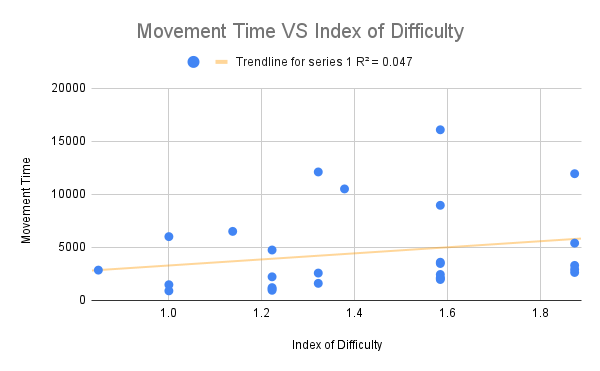}
    \caption{Correlation of movement time with the index of difficulty.}
    \label{fig:mtvsid}
\end{figure}

\subsection{Fitts' Law}
\par Fitts' Law is a fundamental principle in understanding users' sensory-perceptual and perceptual-motor capabilities. The law states that the time required to move toward a target (Movement Time, MT) is a function of the target's width (W) and distance (D) \cite{fitts1954information}. It explores the relationship between the Index of Difficulty (ID) and MT. The ID, as shown in Equation \ref{eq:1}, measures the complexity of a movement task, calculated based on the ratio of movement distance (Amplitude, A) to the target width. MT, as shown in Equation \ref{eq:2}, is a function of the task's ID and is directly proportional to it. We measured the average completion time for each task and used the results as the MT. 

\par Since Fitts' original experiments \cite{fitts1954information} focused on pointing tasks in a 1-dimensional (1D) context, we extended Fitts' Law into a 3D object manipulation context, inspired by the ID equation proposed by Ha and Woo (\citeyear{ha2010empirical}). In our case, the ID formula is represented by Equation \ref{eq:3}, where $O_1$ is the width of the stationary tangible cube and $O_2$ is the width of the movable tangible cube. In addition to ID and MT, we also considered Throughput (TP) to evaluate the performance of PointARs in the context of our primary task. TP, as shown in Equation \ref{eq:4}, is an efficiency measure (measured in bits per second, bps) for human-computer interaction tasks, combining both speed and accuracy to assess the overall performance of an input device or system \cite{douglas1999testing}.
\begin{equation}\label{eq:1}
ID = \log_2(\dfrac{A}{W} + 1)
\end{equation}

\begin{equation}\label{eq:2}
MT = a + b \times ID
\end{equation}

\begin{equation}\label{eq:3}
ID = \log_2(\dfrac{A}{min(O_1, O_2)} + 1)
\end{equation}

\begin{equation}\label{eq:4}
TP = \dfrac{ID}{MT}
\end{equation}

\subsection{Results}
\par Figure \ref{fig:completion-time} summarizes the average completion time of the primary task across varying combinations of tangible cube widths (in inches) and distances (in inches). These results align with Fitts' Law, which suggests that tasks with larger distances between targets take more time to complete. However, contrary to Fitts' Law, which also posits that smaller targets increase task complexity, our findings indicate an inverse relationship between tangible cube size and completion time: larger stationary cubes correspond to longer completion times. This may be due to the challenges of accurately tracking larger cubes within the PointAR system. The size of the movable tangible cube also impacts completion times; for a stationary cube of 1.5 inches, larger movable cubes result in longer completion times. When the stationary cube is 2 inches, the quickest completion time is observed with a movable cube of the same size, but completion times increase when the movable cube is 2.5 inches. Interestingly, a 2-inch movable cube yields the lowest completion time when paired with a 2.5-inch stationary cube, while a smaller 1.5-inch movable cube results in the highest completion times.

\par Figure \ref{fig:tp} shows the Throughput (TP) for PointARs based on varying tangible cube widths and distances. These results are consistent with the formula for TP, where MT is inversely proportional to TP. For example, when both cubes are 1.5 inches and placed 2 inches apart, the average completion time is low, while the corresponding TP is high. 

\par Additionally, our data reveals a trend in line with Fitts' Law, indicating that higher Index of Difficulty (ID) values correlate with increased MT. This supports the law's prediction that more complex tasks require more time to complete. This relationship was further confirmed by a correlation coefficient (R\textsuperscript{2}) of 0.047, indicating a positive linear relationship between ID and MT, despite some overlapping data points in Figure \ref{fig:mtvsid}. The overlap suggests that the ID equation we used may need refinement to better distinguish between the different configurations.

\par Finally, regarding the average error across different configurations, we observed that as the size difference between the stationary and movable cubes increases, so does the average error, as shown in Figure \ref{fig:error}. This could be attributed to the increased complexity of the task when using larger movable cubes.

\subsection{Analysis}
\par Among all the configurations, the one where both the stationary and movable tangible cubes are 2 inches in size and placed 2 inches apart yielded the lowest movement time (MT). In contrast, the configuration with a 2.5-inch movable cube, a 2-inch stationary cube, and a distance of 4 inches between them resulted in the highest MT.

\par The results also indicated that the configuration yielding the highest throughput (TP) involved both the stationary and movable tangible cubes being 1.5 inches in size and placed 2 inches apart. This configuration exhibited the highest efficiency, suggesting that participants could complete the task more quickly and accurately. On the other hand, the configuration with the lowest TP involved a 2.5-inch movable cube, a 2-inch stationary cube, and a distance of 4 inches between them, indicating that this setup was less efficient for users.

\par These results can be explained by the fact that larger movable cubes made the task more challenging and increased the likelihood of errors, especially when compared to the stationary cubes.



\section{Conclusion and Recommendations}
\par In this study, we extended Fitts' Law to a 3D space by evaluating the current PointARs system with a primary task involving 3D object manipulation. For our experimental setup, we designed 9 configurations based on different pairings of tangible cube dimensions and distances, and evaluated them by measuring interaction metrics such as completion time (MT), error rate, and throughput (TP). Our findings align with Fitts' Law in that larger distances between targets increased task complexity. However, we also observed that as the size of the tangible cubes increased, task complexity grew, which contradicts the law's expectation that smaller targets would increase complexity. This may be due to limitations in the PointARs system's ability to track larger cubes. Based on these results, we recommend redesigning the current PointARs tangible cubes to a size range of 1.5 to 2 inches and limiting the distance between cubes to 2 inches when dealing with multiple objects in the system’s projection space. Future studies should explore additional configurations and tangible shapes to further extend Fitts' Law to other 3D object manipulation tasks and refine the evaluation of the PointARs system.

\bibliography{main}

\end{document}